\documentclass[aps,prl,10pt,twocolumn,a4paper,superscriptaddress,  notitlepage]{revtex4-1}

\usepackage{amsmath}
\usepackage{amssymb}
\usepackage{textcomp}
\usepackage{upgreek}
\usepackage{units}

\usepackage{hyperref}
\hypersetup{colorlinks=true, citecolor=blue, urlcolor=blue, linkcolor=blue}

\usepackage[pdftex]{graphicx}

\usepackage{microtype}

\usepackage{txfonts}

\renewcommand{\vec}[1]{\ensuremath{\boldsymbol{#1}}}
\newcommand{\mathds}[1]{\mathbb{#1}}

\begin{document}

\title{
Quantum Damping of Skyrmion Crystal Eigenmodes due to Spontaneous Quasiparticle Decay
}

\author{Alexander Mook}
\affiliation{Department of Physics, University of Basel, Klingelbergstrasse 82, CH-4056 Basel, Switzerland}

\author{Jelena Klinovaja}
\affiliation{Department of Physics, University of Basel, Klingelbergstrasse 82, CH-4056 Basel, Switzerland}

\author{Daniel Loss}
\affiliation{Department of Physics, University of Basel, Klingelbergstrasse 82, CH-4056 Basel, Switzerland}

\begin{abstract}
    The elementary excitations of skyrmion crystals experience both emergent magnetic fields and anharmonic interactions brought about by the topologically nontrivial noncollinear texture. The resulting flat bands cause strong spontaneous quasiparticle decay, dressing the eigenmodes of skyrmion crystals with a finite zero-temperature quantum lifetime. Sweeping the flat bands through the spectrum by changing the magnetic field leads to an externally controllable energy-selective magnon breakdown. In particular, we uncover that the three fundamental modes, i.e., the anticlockwise, breathing, and clockwise mode, exhibit distinct decay behavior, with the clockwise (anticlockwise) mode being the least (most) stable mode out of the three.
\end{abstract}

\maketitle


\paragraph{Introduction.}
Landau's quasiparticle concept is a standard of condensed matter physics textbooks and frequently adopted to describe properties of solid state systems \cite{landau1980statistical}. However, many-body interactions can lead to spontaneous quasiparticle decay (SQD), dressing quasiparticles with intrinsic lifetimes and potentially wiping out their spectral weight. Such a quantum-mechanical many-body phenomenon is known from phonons in crystals \cite{Baumgartner1981, Eisenmenger1984} and liquid helium \cite{pitaevskii1959properties, Zawadowski1972, Graf1974, Smith1977, Glyde1998, Fak1998}, or magnetic excitations in spin liquids \cite{Zhitomirsky2006SL, Stone2006, Doretto2012} and quantum (anti)ferromagnets \cite{Chernyshev2006NonColl, Chernyshev2009, Zhitomirsky2013, Oh2013, Chernyshev2015, Chernyshev2015largeS, Oh2016, Du2015, Du2016, Winter2017, McClarty2019, Rau2019}. Magnetic systems are particularly attractive platforms for SQD, because magnetic fields serve as handles to manipulate both the magnonic dispersion and interactions \cite{Zhitomirsky1999, Syljuasen2008, Mourigal2010, Masuda2010, Stephanovich2011, Chernyshev2012, Fuhrmann2012, Stephanovich2014, Chernyshev2016, Hong2017, McClarty2018}.

\begin{figure}
    \centering
    \includegraphics[width = \columnwidth]{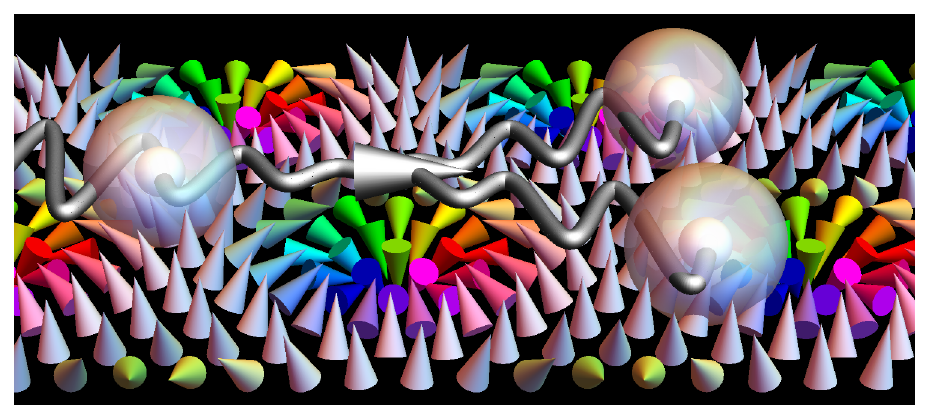}
    \caption{Sketch of spontaneous magnon decay in a skyrmion crystal (SkX) sheet. Colored arrows in the background indicate the magnetic texture of the SkX. White spheres and wavy lines indicate propagating magnons. A magnon incoming from the left spontaneously decays into two other magnons. This number non-conserving process is brought about by the noncollinear texture and gives rise to a zero-temperature quantum damping of the SkX's eigenmodes.}
    \label{fig:artistic}
\end{figure}

In recent years, magnetic skyrmion crystals (SkXs), as depicted in Fig.~\ref{fig:artistic}, have attracted much attention. These arrays of topologically nontrivial magnetic whirls appear in bulk, multilayers, and thin films, at elevated and zero temperature, in metals and insulators \cite{bogdanov1989thermodynamically, Bogdanov2001, Roessler2006, Muhlbauer2009, Yu2010, Munzer2010, Yu2011, Seki2012, Kezsmarki2015, Finocchio2016, Seki2016book, Soumyanarayanan2017, Nayak2017, Kuramaji2017, Fert2017, Jiang2017, Zhou2018, EverschorSitte2018}. Besides numerous fundamental discoveries \cite{Binz2008, Lee2009, Neubauer2009, Schulz2012, Mochizuki2012, Li2013, Hoogdalem2013, Mochizuki2014Ratchet, Mook2017a, Han2019} they spawned visions of future spintronic \cite{Iwasaki2013, Tomasello2014, Kim2019Review}, magnonic \cite{SchutteGarst2014, Ma2015, Ma2015b, Molina2016, Buhl2017, Gobel2017, Zhang2017, Garst2017, Diaz2019, Lin2019, Kim2019, Diaz2019arxiv, Xing2019arxiv}, logic \cite{Zhang2015, Luo2018, Kang2018, Mankalale2019}, and unconventional computation applications \cite{Pinna2018, Zazvorka2019}.
Skyrmions and other magnetic textures may also be used in quantum information \cite{Graf2018, MartnezPrez2019} based on cavity optomagnonics \cite{ViolaKusminskiy2019}. 

However, the necessity of device miniaturization drives skyrmions into the realm of quantum physics, where classical theories, which constitute the overwhelming majority of skyrmion studies, fail. So far, only a handful of publications considered quantum properties of skyrmions \cite{Lin2013, RoldanMolina2015, Takashima2016, Aristov2016, Diaz2016, Psaroudaki2017, Doucot2018, DerrasChouk2018, Psaroudaki2018, Ochoa2018, Sotnikov2018, Psaroudaki2019, Lohani2019, Vlasov2019, Psaroudaki2019Arxiv}.

Herein, we explore the many-body quantum physics of elementary excitations in SkXs, which turn out to be a promising platform to study the fundamental phenomenon of SQD, as  sketched in Fig.~\ref{fig:artistic}. The noncollinear texture of SkXs not only installs anharmonic magnon interactions but also emergent magnetic fields. Consequently, some magnon bands are flat akin to Landau levels, giving rise to sharp peaks in the density of states (DOS) of the two-magnon continuum, into which single magnons decay. Manipulating these peaks by a magnetic field allows for a field-tunable energy-selective magnon breakdown detectable by scattering or absorption experiments.

\paragraph{Theory.}
Interacting spins $\hat{\vec{S}}_{\vec{r}}$, localized at lattice sites with position vector $\vec{r}$, are effectively described by spin Hamiltonians 
$
    \hat{H} 
    = 
    \frac{1}{2} \sum_{ \vec{r}, \vec{r}' } \hat{\vec{S}}_{\vec{r}} \cdot \mathds{I}_{\vec{r}, \vec{r}'} \cdot \hat{\vec{S}}_{\vec{r}'} 
    +
    b \sum_{\vec{r}} \hat{S}^z_{\vec{r}}
$,
with interaction matrices $\mathds{I}_{\vec{r}, \vec{r}'}$ and a Zeeman energy due to magnetic field $b$. The classical ground state of $\hat{H}$, obtained by treating the $\hat{\vec{S}}_{\vec{r}}$'s as classical vectors $\vec{S}_{\vec{r}}$ in $\mathds{R}^3$, defines local reference frames $\{ \vec{e}^x_{\vec{r}}, \vec{e}^y_{\vec{r}}, \vec{e}^z_{\vec{r}} \}$, with $\vec{e}^z_{\vec{r}}$ along the classical ground state direction. Excitations above this ground state are captured by the Holstein-Primakoff transformation \cite{Holstein1940} to bosons $\hat{a}_{\vec{r}}^{(\dagger)}$:
$
    \hat{\vec{S}}_{{\vec{r}}} 
    = 
    \sqrt{S} ( \hat{f}_{{\vec{r}}} \hat{a}_{{\vec{r}}} \vec{e}_{{\vec{r}}}^{-} + \hat{a}_{{\vec{r}}}^\dagger \hat{f}_{{\vec{r}}} \vec{e}_{{\vec{r}}}^{+} ) 
    + ( S - \hat{a}_{{\vec{r}}}^\dagger \hat{a}_{{\vec{r}}} ) \vec{e}_{{\vec{r}}}^{z}
$,
with $\vec{e}_{{\vec{r}}}^{\pm} = (\vec{e}_{{\vec{r}}}^{x} \pm \mathrm{i} \vec{e}_{{\vec{r}}}^{y})/ \sqrt{2}$ and spin length $S$.
Expanding $\hat{f}_{\vec{r}} = [ 1-\hat{a}_{{\vec{r}}}^\dagger \hat{a}_{{\vec{r}}}/(2S) ]^{1/2}$ in powers of $1/S$ leads to $\hat{H} = \sum_{t=0}^\infty \hat{H}_t$, where $\hat{H}_t \propto O(S^{2-t/2})$ includes bosonic operators to the $t$th power. $\hat{H}_0$ gives the classical ground state energy, $\hat{H}_1$ vanishes if the magnetic texture is stable, $\hat{H}_2$ describes free magnons, and $\hat{H}_{t>2}$ interactions. In particular, $\hat{H}_3$ comprises number non-conserving three-magnon interactions.

Often, however, linear spin-wave theory ($\hat{H}_2$) is sufficient, because the single-particle sector is disconnected from many-particle sectors in one of the following limits:
(1) Classical spins $S \to \infty$, because interactions cause at least $1/S$ corrections to the spectrum, (2) low temperatures $T$ that freeze out thermally activated interactions, or (3) large fields $b$ that polarize the magnet and energetically separate the $m$-magnon manifolds, whose energies grow with $mb$ ($m=1,2,\ldots)$.

Intriguingly, neither of the above limits is applicable to SkXs. (1) Flat bands cause decay singularities at twice their energy, only regularized at the same $1/S$ order that introduces damping in the first place. Thus, there is no $1/S$ smallness in the damping and large remnants of quantum effects in effectively classical spin systems are expected, an effect otherwise known from Kagome antiferromagnets \cite{Chernyshev2015, Chernyshev2015largeS}. (2) Three-magnon interactions ($\hat{H}_3$) cause \emph{spontaneous} decay even at $T=0$ \cite{Zhitomirsky2013}. (3) Field polarization destroys the SkX and cannot be used.

This \textit{prima facie} case for stuyding magnon-magnon interactions in SkXs calls for a nonlinear spin-wave theory of chiral magnets. Here, we consider a two-dimensional chiral magnet on the triangular lattice (in the $xy$ plane) with lattice constant $a$. Its interaction matrices with elements 
$
    (\mathds{I}_{\vec{r},\vec{r}'})_{mn} = -J \delta_{mn} + \epsilon_{mnp} D_{\vec{r},\vec{r}'}^p
$
include isotropic symmetric ferromagnetic exchange $J>0$ and antisymmetric exchange denoted by a Dzyaloshinskii-Moriya (DM) \cite{Dzyaloshinsky58, Moriya60} vector $\vec{D}_{\vec{r},\vec{r}'} = D \vec{e}_z \times \vec{e}_{\vec{r},\vec{r}'}$ of length $D$ that complies with interfacial inversion symmetry breaking \cite{FertLevy1980, Crepieux1998}; $\vec{e}_z$ is a unit vector in $z$ direction and $\vec{e}_{\vec{r},\vec{r}'}$ in bond direction.

For $D \ne 0$ and $b = 0$, the ground state spin spiral has a pitch of
$
    \lambda \approx \sqrt{3} \uppi a \mathrm{atan}^{-1} [ \sqrt{3}D/(2J) ]
$
[see Sec.~I A of the Supplemental Material (SM) \cite{Supplement}]. Once $b > b_\mathrm{c,1} \approx 0.2 D^2/J$ \cite{Han2010}, a N\'{e}el SkX forms. A commensurate SkX with $\lambda = M a$ ($M$ integer) and $N=M^2$ spins per magnetic unit cell is obtained for
$
    D/J \approx \tan( 2\uppi/M )
$.
Field polarization is reached at $b>b_\mathrm{c,2} \approx 0.8 D^2/J$ \cite{Han2010}.

Numerically, we proceed as follows. We choose $M$, compute $D$ and $J$, and equilibrate a random spin configuration at $T=0$. We obtain the single-particle energies $\varepsilon_{\vec{k},\nu}$ by diagonalizing $\hat{H}_2$ in terms of normal mode bosons $\hat{b}^{(\dagger)}_{\vec{k},\nu}$, where $\nu = 1,\ldots,N$ is the band index and $\vec{k}$ crystal momentum (see Sec.~I B of SM \cite{Supplement}). The cubic decay processes, within which a magnon in state $(\vec{p},\nu)$ decays into two magnons respectively in states $(\vec{k},\lambda)$ and $(\vec{q},\mu)$, read (see Sec.~I C of SM \cite{Supplement})
\begin{align}
    \hat{H}_3^\mathrm{d} 
    &= 
    \frac{\sqrt{S}}{\sqrt{N_\mathrm{u}}} \sum_{\lambda, \mu, \nu}  
    \sum_{\vec{k}, \vec{q}, \vec{p}}^{\vec{p}=\vec{k}+\vec{q}} \left(
        \frac{1}{2! 1!} \mathcal{V}^{\lambda\mu \leftarrow \nu}_{\vec{k}, \vec{q} \leftarrow \vec{p}}
        \hat{b}^\dagger_{\vec{k}, \lambda}  \hat{b}^\dagger_{\vec{q}, \mu}  \hat{b}_{\vec{p}, \nu} 
        +
        \mathrm{H.c.}
        \right)
        \label{eq:H3_normal_modes}
\end{align}
($N_\mathrm{u}$ number of unit cells). The vertex $\mathcal{V}^{\lambda\mu \leftarrow \nu}_{\vec{k}, \vec{q} \leftarrow \vec{p}}$ comprises the interaction strength and the H.c.~part a magnon coalescence.

To account for $\hat{H}_3$, we perform second-order many-body perturbation theory.
Concentrating on dynamical $1/S$ corrections to the spectrum, we approximate the single-particle Green's function by
$
    G^{-1}_{\vec{k},\nu}(\varepsilon) \approx \varepsilon - \varepsilon_{\vec{k},\nu} + \mathrm{i} \varGamma_{\vec{k},\nu}
$,
from which we obtain the spectral function
$
    A_{\vec{k},\nu}(\varepsilon) \approx - \mathrm{Im} G_{\vec{k},\nu}(\varepsilon)  / \uppi 
$
of band $\nu$. Evaluating single-bubble diagrams within the on-shell approximation, we derive the spontaneous zero-temperature damping (see Sec.~I D of SM \cite{Supplement})
\begin{align}
    \varGamma^\mathrm{spon}_{\vec{k},\nu}
    &=
      \frac{\uppi}{2 N_\mathrm{u}} \sum_{\vec{q}} \sum_{\lambda, \mu}
      \left| \mathcal{V}^{\lambda \mu \leftarrow \nu}_{ \vec{q}, \vec{k}-\vec{q} \leftarrow \vec{k}} \right|^2
       \delta\left(\varepsilon_{\vec{k},\nu} - \varepsilon_{\vec{q},\lambda} - \varepsilon_{\vec{k}-\vec{q},\mu} \right).
     \label{eq:Gamma}
\end{align}


\begin{figure}
    \centering
    \includegraphics[width = \columnwidth]{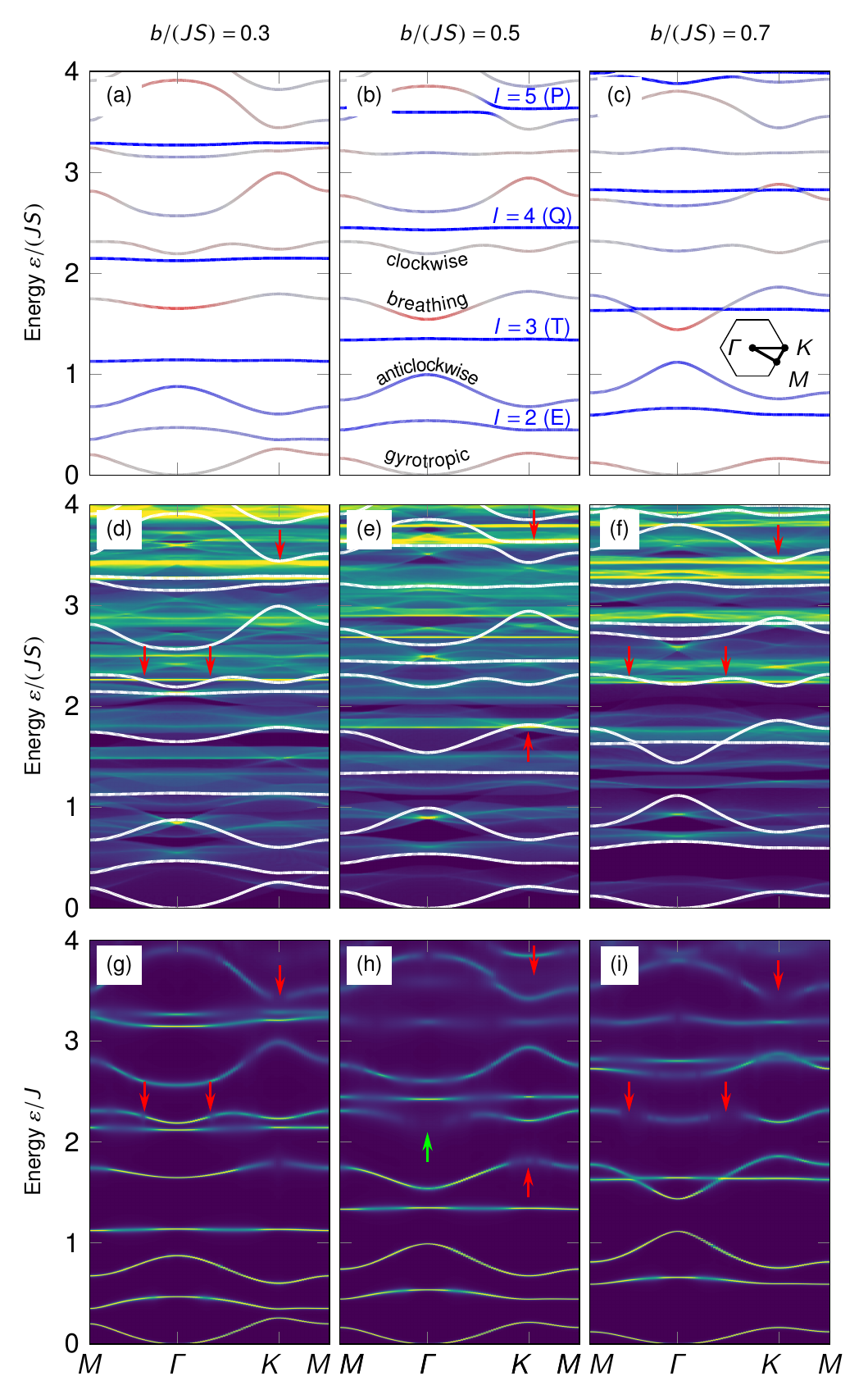}
    \caption{Low-energy portion of the $M=8$ SkX magnon spectrum in dependence on magnetic field $b$. (a)-(c) Single-particle magnon energies $\varepsilon_{\vec{k},\nu}$, whose blue/gray/red color encodes negative/zero/positive magnetic moment $\mu_{\vec{k},\nu}$. (d)-(f) Two-magnon DOS $\mathcal{D}^\mathrm{2m}_{\vec{k}}(\varepsilon)$ (blue: zero; yellow: maximal), with red arrows indicating selected points where $\varepsilon_{\vec{k},\nu}$ (white lines) crosses regions of large $\mathcal{D}^\mathrm{2m}_{\vec{k}}(\varepsilon)$. (g)-(i) Spectral function $A_{\vec{k}}(\varepsilon)$ for $S=1$, with sharp yellow/broad blue quasiparticle peaks indicating undamped/damped magnons. Strong damping is found for the crossing points (red arrows).}
    \label{fig:kDependence}
\end{figure}

\paragraph{Results.}
First, let us establish results of the harmonic theory.
Figures~\ref{fig:kDependence}(a)-(c) show the dispersion of the lowest free-magnon bands $\varepsilon_{\vec{k},\nu}$ of an $M=8$ SkX in dependence on $b$ ($D/J=1$). We identify Landau-level-like flat modes that derive from single-skyrmion modes associated with an $l$th-order polygon deformation in real space \cite{Lin2014} [$l=2$: elliptic mode (E), $l=3$: triangular mode (T), $l=4$: quadrupolar mode (Q), $l=5$: pentagonal mode (P), \dots]. Moreover, there are dispersive modes, the lowest four of which respectively coincide with the gyrotropic (G), anticlockwise (A), breathing (B), and clockwise mode (C) at the Brillouin zone center \cite{Diaz2019arxiv} [cf.~labels in Fig.~\ref{fig:kDependence}(d)]. The G mode derives from the translational $l=1$ mode of a single skyrmion \cite{Petrova2011}. 

The color of the bands in Figs.~\ref{fig:kDependence}(a)-(c) indicates the magnonic out-of-plane magnetic moment
$
    \mu_{\vec{k},\nu} 
    = - \partial \varepsilon_{\vec{k},\nu} / \partial b
$.
A positive (negative) $\mu_{\vec{k},\nu}$ means that an increasing field shifts the modes towards lower (higher) energies. Flat modes carry large negative $\mu_{\vec{k},\nu}$, because of additional angular momentum associated with $l$. Hence, upon increasing the field, they overtake the dispersive modes, in agreement with Ref.~\onlinecite{Diaz2019arxiv}. 

We exploit this field tunability to engineer SQD. To appreciate this idea, note that $\varGamma_{\vec{k},\nu}^\mathrm{spon}$ is a weighted two-magnon DOS
$
    \mathcal{D}^\mathrm{2m}_{\vec{k}}(\varepsilon) = \frac{1}{N_\mathrm{u}} \sum_{\vec{q}} \sum_{\lambda, \mu}
    \delta (\varepsilon - \varepsilon_{\vec{q},\lambda} - \varepsilon_{\vec{k}-\vec{q},\mu} )
$
at $\varepsilon = \varepsilon_{\vec{k},\nu}$. In Figs.~\ref{fig:kDependence}(d)-(f), the $\varepsilon_{\vec{k},\nu}$'s are overlaid with color maps that show $\mathcal{D}^\mathrm{2m}_{\vec{k}}(\varepsilon)$. Besides extended regions of moderate $\mathcal{D}^\mathrm{2m}_{\vec{k}}(\varepsilon)$, several horizontal features, i.e., sharp yellow lines, at twice the energy (or at the sum of two energies) of flat modes are identified. If a single-magnon branch crosses a region of large $\mathcal{D}^\mathrm{2m}_{\vec{k}}(\varepsilon)$, it is ``in resonance'' with flat modes and the kinematically allowed phase space for SQD, i.e., the number of decay channels that fulfill both energy and momentum conservation, is particularly large. Selected instances of such crossings are marked by red arrows and pronounced $\varGamma_{\vec{k},\nu}^\mathrm{spon}$ is expected. 

We confirm this prediction by calculating $\varGamma^\mathrm{spon}_{\vec{k},\nu}$ and $A_{\vec{k}}(\varepsilon) = \sum_{\nu=1}^N A_{\vec{k},\nu}(\varepsilon)$, the latter of which is shown in Figs.~\ref{fig:kDependence}(g)-(i). Stable magnons appear as sharp yellow quasiparticle peaks, while strongly damped magnons experience considerable lifetime broadening (broad blue features). For example, consider the regions of pronounced damping marked by red arrows in Fig.~\ref{fig:kDependence}(g), which can be traced back to $\mathcal{D}^\mathrm{2m}_{\vec{k}}(\varepsilon)$ in Fig.~\ref{fig:kDependence}(d). 
These flat-band resonances can be so large that the spectral weight is almost completely wiped out, as exemplified by the pentagonal $l=5$ mode at $b/(JS) = 0.5$ in Figs.~\ref{fig:kDependence}(b), (e) and (h); cf.~upper right red arrow. However, a slight magnetic-field detuning of the resonance condition leads to a reappearing quasiparticle peak. To confirm so, see Sec.~II A of SM \cite{Supplement}.

Besides flat-mode resonances, one identifies several regions with large damping but moderate $\mathcal{D}^\mathrm{2m}_{\vec{k}}(\varepsilon)$, associated with decays into (at least) one dispersive mode. As an example, consider the C mode in Fig.~\ref{fig:kDependence}(h), indicated by the green arrow. Its strong damping is not related to a particularly large $\mathcal{D}^\mathrm{2m}_{\vec{k}}(\varepsilon)$ in Fig.~\ref{fig:kDependence}(e). This mechanism is the prime cause of high-energy magnon damping in Figs.~\ref{fig:kDependence}(g)-(i), reflecting the abundance of kinematically allowed decay channels. 

Since all decay channels that contain at least one flat mode as decay product exhibit a strong field dependence, the SQD is field tunable and highly energy selective, which sets SkXs apart from other SQD platforms. This finding can be confirmed by inelastic neutron or resonant X-ray scattering experiments, which measure the dynamical structure factor, an object directly related to the spectral function \cite{Lovesey1977}. A rich magnetic-field dependence of measured line widths is expected. To pursue this thought further, we continue with an analysis of uniform excitations ($\vec{k}=\vec{0}$), which can even be studied by absorption experiments \cite{Onose2012, Okamura2013, Schwarze2015, Ehlers2016}. For the geometry studied here, i.e., with magnetic field normal to the SkX, only the ABC modes are active \cite{Mochizuki2012}. However, inclined magnetic fields admit  additional active modes \cite{Ikka2018}.

\begin{figure}
    \centering
    \includegraphics[width = \columnwidth]{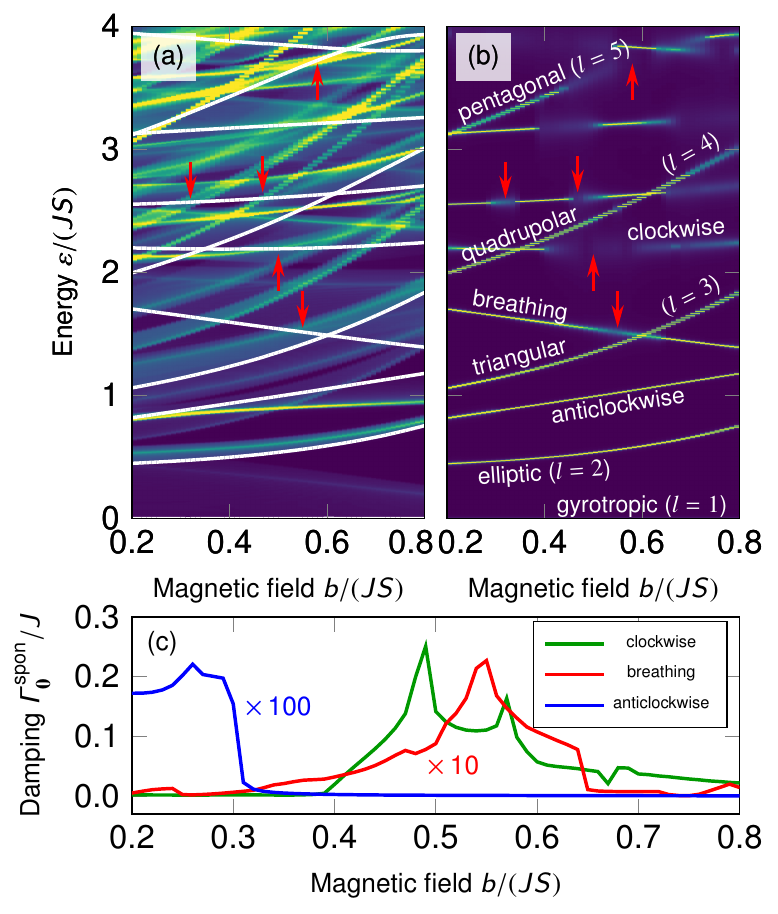}
    \caption{Magnon decay at the Brillouin zone center ($\vec{k}=\vec{0}$) of an $M=8$ SkX in dependence on magnetic field $b$. 
	(a) Two-magnon DOS $\mathcal{D}^\mathrm{2m}_{\vec{0}}(\varepsilon)$ (blue: zero; yellow: maximal), with red arrows indicating selected points where the eigenmode energies $\varepsilon_{\vec{0},\nu}$ (white lines) cross regions of large $\mathcal{D}^\mathrm{2m}_{\vec{0}}(\varepsilon)$. 
    (b) Spectral function $A_{\vec{0}}(\varepsilon)$ for $S=1$, with sharp yellow/broad blue quasiparticle peaks indicating undamped/damped magnons. Strong damping is found for the crossing points (red arrows). (c) The spontaneous damping $\varGamma^\mathrm{spon}_{\vec{0},\nu}$ of the three magnetically active modes reveals that the clockwise (anticlockwise) mode is the least (most) stable.}
    \label{fig:atGamma}
\end{figure}

\begin{figure*}
    \centering
    \includegraphics[width = \textwidth]{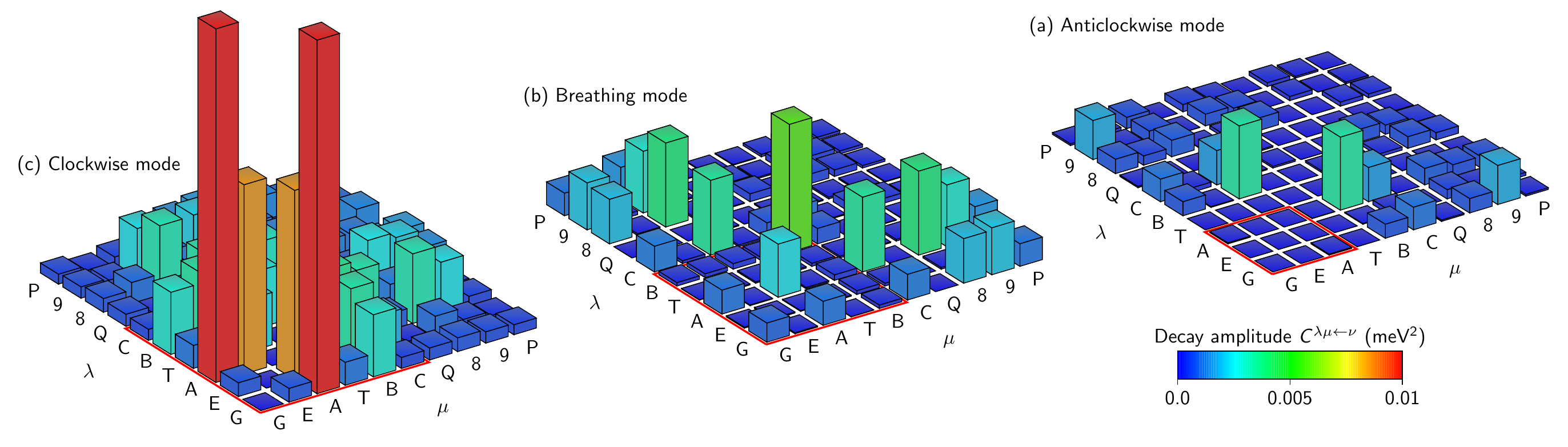}
    \caption{Decay channel analysis of an $M=8$ SkX at $b/(JS)=0.5$ and $\vec{k} = \vec{0}$. Histograms show the decay amplitude $C^{\lambda \mu \leftarrow \nu}$ for $\lambda,\mu = 1, \ldots, 10$ and (a) $\nu = 3$ anticlockwise mode (A), (b) $\nu=5$ breathing mode (B), and (c) $\nu=6$ clockwise mode (C). The histograms are symmetric due to decay product symmetry ($\lambda \leftrightarrow \mu$). Red squares indicate the subsection of decay channels that could, in principle, fulfill energy conservation. By comparing the amplitudes of these channels between the three modes, the clockwise (anticlockwise) mode is found to interact strongly (weakly) with other modes. This explains the two orders of magnitude difference in the damping [cf.~Fig.~\ref{fig:atGamma}(c)].}
    \label{fig:Histogram}
\end{figure*}

Figure~\ref{fig:atGamma}(a) shows $\mathcal{D}^\mathrm{2m}_{\vec{0}}(\varepsilon)$ in dependence on $b$. Applying the same logic as before, strong damping is found when $\varepsilon_{\vec{0},\nu}$ crosses a region of large $\mathcal{D}^\mathrm{2m}_{\vec{0}}(\varepsilon)$. Several of such encounters are highlighted by red arrows, from which a rich structure of strong quasiparticle peak broadening is expected. Indeed, $A_{\vec{0}}(\varepsilon)$ in Fig.~\ref{fig:atGamma}(b) reveals that high-energy magnons exhibit field intervals within which their quasiparticle peak disappears and reappears. In contrast, low-energy magnons exhibit negligible quantum damping, in agreement with what is known for single metastable skyrmion excitations in achiral ferromagnets \cite{Aristov2016}. This suggests long lifetimes of chiral edge magnons between the topologically nontrivial A and B modes \cite{Molina2016, Diaz2019arxiv}.

There are considerable differences in damping between the ABC modes [cf.~Fig.~\ref{fig:atGamma}(c)]. SQD is particularly strong for the C mode, whose damping is as large as $0.2 J$, rendering it on par with frustrated quantum antiferromagnets \cite{Chernyshev2006NonColl, Chernyshev2009, Zhitomirsky2013}.
In contrast, the damping of the B (A) mode is a factor of ten (hundred) smaller. These differences cannot solely be explained in terms of $\mathcal{D}^\mathrm{2m}_{\vec{0}}(\varepsilon)$ but must be looked for in $\mathcal{V}^{\lambda\mu \leftarrow \nu}_{\vec{k}, \vec{q} \leftarrow \vec{p}}$.

We analyze each decay channel $\nu \to (\lambda,\mu)$ by measuring its integrated decay amplitude 
$
    C^{\lambda \mu \leftarrow \nu} \equiv \frac{1}{N_\mathrm{u}} \sum_{\vec{q}} 
    | \mathcal{V}^{\lambda \mu \leftarrow \nu}_{ \vec{q}, -\vec{q} \leftarrow \vec{0}} |^2
$.
It encodes dynamical rather than kinematic details. Results for $1000$ decay channels, involving the ten lowest modes at $b/(JS)=0.5$, are presented in Sec.~II B of SM \cite{Supplement}. Here, we focus on the decay of ABC modes (Fig.~\ref{fig:Histogram}).

The A mode [$\nu=3$, Fig.~\ref{fig:Histogram}(a)] exhibits small interaction with other modes, especially the G und E modes, i.e., those modes that could possibly fulfill energy conservation. Thus, the A mode is the most stable out of the ABC modes for both kinematic as well as dynamic reasons. This explains its negligible damping observed in Fig.~\ref{fig:atGamma}.

The B mode [$\nu=5$, Fig.~\ref{fig:Histogram}(b)] prefers channel $5 \rightarrow (5,5)$, i.e., $\mathrm{B} \rightarrow (\mathrm{B},\mathrm{B})$---which, however, can never obey energy conservation---and channels that involve the G and/or A mode. These channels cause the moderate damping of the B mode in Fig.~\ref{fig:atGamma}. In contrast, channels with polygon modes E, T, and Q as decay products are suppressed by two orders of magnitude.

Finally, being highest in energy, the C mode [$\nu=6$, Fig.~\ref{fig:Histogram}(c)] has several kinematic possibilities to decay. On top of that come pronounced instabilities towards the A and B mode. Thus, the C mode is the least stable among the ABC modes, complying with strong damping (Fig.~\ref{fig:atGamma}).

These findings are corroborated by the nontrivial $b$ dependence of the ABC modes' interaction strengths, presented in Sec.~II C of SM \cite{Supplement}. While that of the C (and B mode) increases as the field increases, that of the A mode decreases. 

Polygon modes have overall larger interaction amplitudes than the ABC modes (Sec.~II B of SM \cite{Supplement}). Decay processes that conserve azimuthal number $l$, e.g., a decay of the P into T and E modes ($l=5=3+2$), exhibit particularly large amplitudes. Recall that isolated skyrmions possess continuous rotation symmetry that conserves $l$, strictly ruling out nonconserving decay channels \cite{Schultheiss2019}. However, the hexagonal deformation in a SkX breaks rotation symmetry and $l$ is no longer conserved. Thus, the preference for $l$-conserving decay channels is a remnant of an isolated skyrmion's symmetry.

The results discussed so far apply to zero temperature. Finite temperatures admit of additional damping, because collision channels that require thermally excited magnons open up. Thus, on top of damping due to spontaneous down conversion comes that due thermally activated down and up conversion. In Sec.~II E of SM \cite{Supplement}, we show that at $k_\mathrm{B} T/J = 0.2$ and $b/JS = 0.485$ ($k_\mathrm{B}$ Boltzmann constant) the thermal damping of the C mode is still only half of its quantum damping. Thus, even the quantum dynamics of SkXs stabilized by moderate thermal fluctuations may be dominated by SQD \footnote{However, for consistency of spin-wave theory, $k_\mathrm{B} T \ll JS$ must hold (see Sec.~II F of SM \cite{Supplement}).}.


\paragraph{Discussion and Conclusion.}
Spontaneously decaying magnons cause field-tunable intrinsic zero-temperature  damping of SkX eigenmodes. This damping is a quantum correction to the classical damping $\varGamma^\alpha_{\vec{k},\nu} \approx \alpha \varepsilon_{\vec{k},\nu} \propto O(S)$ due to phenomenological Gilbert damping $\alpha$.
For the C mode, with $\varepsilon_{\mathrm{C}} \approx J S (D/J)^2$ and $\varGamma^\mathrm{spon}_{\mathrm{C}} \approx 0.1 \times J (D/J)^4$ (see Sec.~II D of SM \cite{Supplement}), the relative importance of the two dampings reads 
\begin{align}
    R_\mathrm{C} \equiv \frac{ \varGamma^\mathrm{spon}_{\mathrm{C}} }{ \varGamma^\alpha_{\mathrm{C}} } 
    \approx
    \frac{0.1}{\alpha S} \left( \frac{D}{J} \right)^2.
    \label{eq:GammaRatio}
\end{align}
Hence, the ideal zero-temperature SkXs to study SQD features small $S$ and $\alpha$ but large $D/J$. Gilbert damping can be as small as $10^{-4}$--$10^{-3}$ both in metals \cite{Mizukami2009, Mankovsky2013, Durrenfeld2015, Andrieu2016, Schoen2016, Husain2016, Lee2017, Qin2017, Shaw2018, bainsla2018low} and electrically insulating skyrmion-hosting compounds like Cu$_2$OSeO$_3$ \cite{Stasinopoulos2017}. Thus, for $S=1$, and $\alpha=10^{-4}$--$10^{-2}$ a ratio $D/J \approx 0.03$--$0.3$ is needed to render $R_\mathrm{C} \approx 1$. Assuming $a = \unit[0.5]{nm}$, this corresponds to a spiral period of $\unit[10]{nm}$--$\unit[100]{nm}$. (This argument ignores inhomogeneous classical damping due to grains \cite{McMichael2003}.)

For thin films of Cu$_2$OSeO$_3$ and MnSi, we respectively estimate $\varGamma^\mathrm{spon}_{\mathrm{C}} \approx \unit[5.6]{neV}$ (using $J/k_\mathrm{B} \approx \unit[50]{K}$ and $D/J \approx 0.06$ \cite{Seki2012}) and $\varGamma^\mathrm{spon}_{\mathrm{C}} \approx \unit[0.1]{\upmu eV}$ ($J/k_\mathrm{B} \approx \unit[29]{K}$ \cite{Muhlbauer2009} and $D/J \approx 0.15$ \cite{Meynell2017}). In absolute terms, these very small values suggest that quantum corrections to the SkX dynamics are negligible in these materials. Therefore, they are suitable for ``topological magnonics'' applications \cite{Molina2016, Diaz2019arxiv}. However, in relative terms, we find $R_\mathrm{C} \approx 3.6$ in Cu$_2$OSeO$_3$ ($S=1$ and $\alpha = 10^{-4}$ \cite{Stasinopoulos2017}), suggesting that quantum damping is larger than classical damping in the low-temperature skyrmion phase \cite{Halder2018}.

The dynamics of the nano-SkX ground state with skyrmion distances of $\unit[1]{nm}$ in Fe/Ir$(111)$ \cite{vonBergmann2006, Grenz2017} is expected to exhibit strong quantum corrections. Although this SkX is stabilized by additional four-spin interactions \cite{Heinze2011} and exchange frustration \cite{Okubo2012, Kamiya2014, Leonov2015, vonMalottki2017, Hayami2019}, which come on top of the DM interaction, we expect $\varGamma^\mathrm{spon}_{\mathrm{C}} > 0.1 J$ simply judging from the tiny skyrmion size.
As far as skyrmions in van der Waals magnets are concerned \cite{Tong2018, Ding2019, Behera2019}, Janus monolayers with large $D/J$ ratios \cite{Liang2019arxiv, Yuan2019arxiv} are promising candidates for strong SQD.
Finally, we also expect strong quantum corrections to the collective dynamics of three-sublattice antiferromagnetic SkXs \cite{Rosales2015, Diaz2019} or $\mathbb{Z}_2$ vortex crystals \cite{Li2019Z2}. In contrast to ferromagnetic SkXs with a locally almost collinear spin texture, their local magnetic texture stays frustrated even in the limit of large topological objects, which adds to the three-magnon vertex.


\acknowledgements
We thank Sebasti\'{a}n D\'{\i}az for helpful discussions. This work was supported by the Georg H.~Endress Foundation and the Swiss National Science Foundation and NCCR QSIT. This project received funding from the European Union’s Horizon 2020 research and innovation program (ERC Starting Grant, grant agreement No 757725).


\bibliography{short,newrefs}

\end{document}